\documentclass[aps,amsmath,amssymb,pra,showpacs]{revtex4}

\usepackage{graphicx}

\begin{document}

\preprint{draft}

\title{Normal and Lateral Casimir Forces between Deformed Plates}

\author{Thorsten Emig}
\affiliation{Institut f\"ur Theoretische Physik, Universit\"at zu
K\"oln, Z\"ulpicher Stra\ss e 77, D-50937 K\"oln, Germany}

\author{Andreas Hanke}
\affiliation{Institut f\"ur Theoretische Physik, Universit\"at Stuttgart,
Pfaffenwaldring 57, D-70550 Stuttgart, Germany}

\author{Ramin Golestanian}
\affiliation{Institute for Advanced Studies in Basic Sciences,
Zanjan 45195-159, Iran \\
and Institute for Studies in Theoretical Physics and Mathematics,
P.O. Box 19395-5531, Tehran, Iran}

\author{Mehran Kardar}
\affiliation{Physics Department, Massachusetts Institute of
Technology, Cambridge, MA 02139}

\date{\today}

\begin{abstract}
  The Casimir force between macroscopic bodies depends strongly on
  their shape and orientation. To study this geometry dependence in
  the case of two deformed metal plates, we use a path integral
  quantization of the electromagnetic field which properly treats
  the many-body nature of the interaction, going beyond the commonly
  used pairwise summation (PWS) of van der Waals forces. For arbitrary
  deformations we provide an analytical result for the
  deformation induced change in Casimir energy, which is exact to second
  order in the deformation amplitude. For the specific case of
  sinusoidally corrugated plates, we calculate both the normal and the
  lateral Casimir forces.  The deformation induced change in the
  Casimir interaction of a flat and a corrugated plate shows an
  interesting crossover as a function of the ratio of the mean plate
  distance $H$ to the corrugation length $\lambda$: For $\lambda \ll
  H$ we find a {\em slower} decay $\sim H^{-4}$, compared to the
  $H^{-5}$ behavior predicted by PWS which we show to be valid only
  for $\lambda \gg H$. The amplitude of the lateral force between two
  corrugated plates which are out of registry is shown to
  have a maximum at an optimal wavelength of $\lambda \approx 2.5 \,
  H$. With increasing $H/\lambda \gtrsim 0.3$ the PWS approach becomes
  a progressively worse description of the lateral
  force due to many-body effects. These results may be of relevance
  for the design and operation of novel microelectromechanical systems
  (MEMS) and other nanoscale devices.
\end{abstract}


\maketitle

\section{Introduction}
\label{section_intro}

More than five decades ago, Casimir predicted that the
ground state energy of photons is alternated in the presence of two
parallel perfectly conducting metal plates in such a way as to lead to
an observable macroscopic force between them
\cite{Casimir48}. The attractive force (per plate area
$A$) has an universal amplitude, an energy scale set by the
fundamental constant $\hbar c$, and decays
with the distance $H$ between the plates as
\begin{equation} \label{flat}
\frac{F}{A} \, = \, - \, \frac{\pi^2}{240} \frac{\hbar c}{H^4} \, \, .
\end{equation}
This remarkable prediction of quantum electrodynamics
has implications in many contexts ranging from surface physics
\cite{Israelachvili92}, particle physics \cite{Milton80}, to
cosmology \cite{Bytsenko+96}. Because of its fundamental nature, the
Casimir effect has motivated extensive theoretical work, especially
during the last decade. The pioneering result in Eq.~(\ref{flat})
has been generalized to include important effects such as the finite
conductivity and surface roughness of the plates, finite temperature,
and even moving plates in the dynamic counterpart to the Casimir effect
(see Refs.~\cite{PMG86,ER91,Milonni94,MT97,Lam99,Kardar99,Bordag+01}
for reviews).

On the experimental front, the early attempts at observing the Casimir
force, initiated by Sparnaay in 1958 \cite{Sparnaay58} and later by
Van Blokland and Overbeek in 1978 \cite{BO78}, were not conclusive due
to large experimental uncertainty.  In recent years, however, there
have been a number of precision measurements which set the modern
stage in this field; starting in 1997 by Lamoreaux \cite{Lamoreaux97}
who used a torsion pendulum with an electromechanical feedback system
to measure the Casimir force between a spherical surface (lens) and a
flat plate.  Mohideen {\em et al.\/} \cite{MR98} measured the Casimir
force between a sphere mounted on the tip of a flexible cantilever and
a flat plate by an atomic force microscope.  Chan {\em et al.\/}
\cite{CAKBC2001} measured the Casimir force between a sphere and a
flat plate in a microelectromechanical system (MEMS) using a
micromachined torsional device. All these experiments
confirm the Casimir force formula in the range from 100\,nm to several
$\mu$m to a few per cent accuracy.  In order to achieve this high
precision, a careful analysis of the corrections due to the finite
conductivity of the metal surfaces, roughness, and nonzero
temperature, is indispensable
\cite{Bordag+01,KNS87,NSC,corr99,LR2000,corr00}.
The above cited
experiments in fact deviate from the flat plate geometry corresponding
to Eq.\,(\ref{flat}) by using a plane-sphere configuration (thus
avoiding the experimental difficulty of keeping two flat surfaces
sufficiently parallel). The force for a spherical surface with (large)
radius $R$ at a distance $H$ of closest surface-to-surface approach
from a flat plate can then be calculated from the Casimir potential
${\cal E}(H)$ for two flat plates by using the {\em proximity force rule}
\cite{D34}, see Eq.~(\ref{F-DA}) below.  Recently, however, G. Bressi
{\em et al.\/} \cite{Bressi02} measured the Casimir force between two
parallel flat surfaces directly, confirming Eq.\,(\ref{flat}) to
15\,\% accuracy.

The Casimir force in Eq.\,(\ref{flat}) has analogies to the effective
force between particles or plates immersed in a system close to its
{\em critical point\/}, which arises due to the modification of
thermal fluctuations of the bulk order parameter. This effect was
originally predicted in 1978 by Fisher and de Gennes \cite{FdG78}
for colloidal particles immersed in a binary liquid
mixture near its critical de-mixing point, and observed experimentally
for silica spheres immersed in a mixture of water and oil
(2,6-lutidine) \cite{BE}.  Related phenomena occur in liquid
crystals \cite{LC}, microemulsions \cite{U01}, and for inclusions in
fluctuating membranes \cite{incl} (see Ref.\,\cite{Kardar99} for a
review).  In recent years, the critical-point Casimir effect has
attracted increasing theoretical
\cite{Kardar99,K91,Krech,G99,HSED98,BAFG02} and experimental interest
\cite{ML99,GC}.  For He$^4$ wetting films close to the superfluid
phase transition, the theoretical predictions \cite{K91} for
critical-point Casimir forces between parallel surfaces exhibiting
Dirichlet boundary conditions have been confirmed quantitatively
\cite{GC}.

Equation (\ref{flat}) for the electromagnetic Casimir force is valid
in the ideal limit of perfectly conducting plates.  In the more
general context of the Lifshitz theory for dielectric bodies
\cite{Lifshitz}, this corresponds to an infinite dielectric constant
$\varepsilon$ for all frequencies $\omega$.  For finite $\varepsilon =
\varepsilon(\omega)$, this power law for the force is recovered for
large distances $H \gg c/\omega_0$, where $\omega_0$ is the smallest
resonance (absorption) frequency of the dielectric (usually
$c/\omega_0\approx 10 \text{--} 100$ nm). In this, so-called retarded,
limit, the force is {\em universal\/} in the sense that it only
depends on the electrostatic dielectric constant $\varepsilon_0 =
\varepsilon(0)$.  The opposite limit of $H \ll c/\omega_0$ gives the
unretarded van der Waals force $F/A \sim H^{-3}$.  The interpretation
of the Casimir force in terms of changes in zero point vacuum
electromagnetic energy suggests it to be a strong function of
geometry; probing the global shape of the boundary that confines the
vacuum fluctuations \cite{source}.  Indeed, whereas the van der Waals
force between electrically polarizable particles is always attractive,
even the {\em sign\/} of the Casimir force is geometry dependent, and
can be {\em repulsive\/}, e.g., for a thin spherical or cubic
conducting shell \cite{Boyer68,Balian,Maclay00,Jaffe02}.  Repulsive
Casimir forces are expected also when {\em magnetic\/} properties
of the boundaries are exploited, e.g., by using a perfectly conducting 
and an infinitely permeable plate \cite{mag}.

Apart from the importance of these phenomena to basic science, the
ongoing refinement of nanofabrication technology in electronics and
mechanics also provides new impetus for the understanding of such
systems in view of applications in nanotechnology.  On length scales
of about 100 nm and below, the Casimir force becomes comparable or
even dominant to other forces \cite{Srivastava+85}, and thus must be
taken into account in the design and operation of nanoscale
devices \cite{Serry+95,Buks+01}.  Indeed, the experiment of Chan {\em
et al.\/} \cite{CAKBC2001} demonstrates the possibility for novel
actuation schemes in microelectromechanical systems (MEMS) based on
the Casimir force.  They also show that the Casimir force can be used
to control {\em dynamic\/} properties of such systems, e.g., in a
nonlinear micromechanical Casimir oscillator \cite{chanosc}.  The
above mentioned dependence of strength and sign of the Casimir force
on geometry and material properties offers the opportunity to
manipulate this interaction in a controlled way, e.g., by tailoring
the shape of the interacting surfaces.  On the other hand, movable
elements of nanoscale devices may unwantedly stick together due to the
strong attractive Casimir force, impeding their operation
\cite{Serry+95,Buks+01}.  This so-called stiction could possibly be prevented
by using shapes (e.g., suitable modulations) and materials of movable
elements such that the Casimir force between them is repulsive (or at least reduced).

Due to the importance of the Casimir force to basic and applied
science, it is highly desirable to demonstrate its strong shape
dependence in a set-up that clearly shows its distinction from the
usual pairwise additive interactions \cite{GK97}.  In a previous
Letter \cite{EHGK01} we pointed out that a promising route to this
end is via modifications of the parallel plate geometry, since
measurement of the putative repulsive Casimir interaction for a
conducting sphere is experimentally difficult.  In searching for
nontrivial boundary dependences, Roy and Mohideen \cite{RM99}
examined the force between a sphere and a sinusoidally corrugated
plate with amplitude $a\approx 60 \, \mbox{nm}$ and wavelength
$\lambda\approx 1.1 \, \mu\mbox{m}$. (This geometry was first
suggested in Refs. \cite{GK97}.) Over the range of separations
$H\approx 0.1 - 0.9 \, \mu\mbox{m}$, the observed force showed
clear deviations from the dependence expected on the basis of
decomposing the Casimir force to a sum of pairwise contributions
(in effect, an average over the variations in separations).  This
experimental result motivated our calculation of the exact Casimir
force in the geometry depicted in Fig.\,\ref{fig1}, without the
assumption of pairwise additivity.  Our analytic results [see
Eq.\,(\ref{ce+}) and Fig.\,\ref{fig2}] hold to second order in
$a$, and show that for fixed $H$ the corrections due to
corrugation strongly depend on $\lambda$.  In fact, for
$H/\lambda\gg 1$ the correction is by a factor of $H/\lambda$ {\em
larger} than in the opposite limit of $H/\lambda\ll 1$ where the
assumption of pairwise additivity is asymptotically correct.
However, the experiments of Ref.\,\cite{RM99} are performed in the
range of $H/\lambda\approx 0.1 - 0.8$ where the corrections to
pairwise additivity may not be significant enough to account for
the observed deviations.  In Ref.\,\cite{KZC2000} it has been
suggested that these deviations are due to a {\em lateral\/} force
that tends to preferentially position the spherical AFM tip on top
of local maxima of the modulated surface (leading to a smaller
separation and stronger force). Based on our results, we thus
propose that the shape dependence of the Casimir force can be
probed more clearly by going to modulations of shorter wavelength;
a hard but achievable goal.

The originally predicted Casimir force between two flat metal
plates (or between a flat and a deformed plate) is, for symmetry
reasons, oriented normal to the surfaces of the plates. However,
if both plates are deformed there is also a lateral Casimir force,
as predicted in Refs. \cite{GK97} and confirmed experimentally
\cite{Chen+02}. To date, the lateral force between two corrugated
plates has been calculated explicitely only within the
approximative approach of a pairwise summation of van der Waals
forces, see, e.g., \cite{Bordag+01}. Here we calculate the lateral
force exactly to second order in the corrugation amplitude for the
geometry shown in Fig.~\ref{fig4}, without referring to a pairwise
summation scheme. As for the normal force, our results [cf.
Eq.~(\ref{Ecc}) and Fig. \ref{fig5}] show that the lateral force
strongly depends on the ratio of the corrugation length $\lambda$
and mean plate separation $H$. We find the pairwise summation to
be a valid approximation only for sufficiently small values of
$H/\lambda \lesssim 0.3$. However, the experiment of Ref.
\cite{Chen+02} is performed at a ratio $H/\lambda \approx 0.18$
where we do not expect significant deviations from the pairwise
summation approximation.

The outline of the paper is as follows: In the next section we
set up the general path integral formulation for the Casimir energy
of the electromagnetic gauge field. By separating into transversal
electric and magnetic modes, we reduce the problem
to two decoupled problems
for scalar fields which differ only in their  boundary
conditions. In each case, we calculate the Casimir energy
for general plate deformations perturbatively. In the third section, 
we give a brief
summary of the pairwise summation approach, and the resulting Casimir
interaction. Section four gives detailed results on both the normal
and the lateral force between sinusoidally corrugated plates. 
We conclude by discussing the relevance of our results to experiments. 
Details of the calculations are left to the appendices.

\section{Path integral formulation of the Casimir energy}
\label{section_path}

We consider two perfectly conducting deformed plates $S_\alpha$
($\alpha = 1, 2$) of mean separation $H$, which are infinitely
extended along the plane spanned by ${\bf y}_{\parallel} = (y_1,
y_2)$. Assuming static and uniaxial deformations without
overhangs, their profiles are described by height functions
$h_{\alpha}(y_1)$, with $\int d y_1 \, h_{\alpha}(y_1) = 0$.  The
Casimir energy at zero temperature corresponds to the difference
of the ground state energies of the quantized electromagnetic (EM)
field for plates at distance $H$ and at $H \to \infty$,
respectively.  To obtain this energy, we employ the path integral
quantization method.  For general deformations, it is necessary to
consider the action
\begin{equation} \label{em-action}
S_{\text{em}}\{A_{\mu}\} \, = \, - \frac{1}{4} \int d^4 X \,
F_{\mu \nu} F^{\mu \nu} \, \, ,
\end{equation}
where $X$ denotes a point of 4D spacetime, $F_{\mu \nu} =
\partial_{\mu} A_{\nu} - \partial_{\nu} A_{\mu}$, and the four
potential $A_{\mu}$ is subject to the boundary condition that the
tangential components of the electric field vanish on the
surfaces. The redundant degrees of freedom due to the gauge
invariance of the electromagnetic field can be eliminated by the
Faddeev-Popov gauge fixing procedure \cite{PS95,GK97}.

However, for the uniaxial deformations under consideration here,
we can develop a simpler quantization scheme, by a similar
reasoning as used in the context of waveguides with constant
cross-sectional shape \cite{EHGK01}.
In this case, the transverse magnetic (TM)
waves and transverse electric (TE) waves (with respect to the
translational invariant $y_2$ direction) constitute a complete set
of modes to describe an arbitrary electromagnetic field between
the plates \cite{Jackson}. For TM waves all field components are
then uniquely given by a scalar function corresponding to the
electric field along the invariant direction,
\begin{equation} \label{TM}
\Phi_{\text{TM}}(t, y_1, y_2, z) = E_2(t, y_1, y_2, z),
\end{equation}
with the Dirichlet boundary condition
$\Phi_{\text{TM}}|_{S_\alpha} = 0$ on each surface $S_\alpha$.
The TE waves are analogously described by the scalar function
\begin{equation} \label{TE}
\Phi_{\text{TE}}(t, y_1, y_2, z) = B_2(t, y_1, y_2, z),
\end{equation}
with the Neumann boundary condition $\partial_n
\Phi_{\text{TE}}|_{S_\alpha} = 0$, where $\partial_n$ is the
normal derivative of the surface $S_\alpha$ pointing into the
space between the two plates.  After a Wick rotation to the
imaginary time variable $X^0 = i c t$, both fields
$\Phi_{\text{TM}}$ and $\Phi_{\text{TE}}$ can be quantized using
the Euclidean action
\begin{equation} \label{e-action}
S_{\text{E}}\{\Phi\} \, = \, \frac{1}{2} \int d^4 X \, (\nabla
\Phi)^2 \, \, .
\end{equation}
In the 4D Euclidean space, the plates are parameterized by
$X_1({\bf
  y}) = [{\bf y}, h_1(y_1)]$ and $X_2({\bf y}) = [{\bf y}, H +
h_2(y_1)]$, where ${\bf y} = (y_0, y_1, y_2) = (y_0, {\bf
  y}_{\parallel})$, and $y_0 = i c t$.

In order to obtain the ground state energy from this
quantization scheme, we now consider the partition functions
${\cal
  Z}_{\text{D}}$ and ${\cal Z}_{\text{N}}$ for the scalar field
Euclidean action $S_{\text{E}}$ both with Dirichlet (D) and
Neumann (N) boundary conditions at the surfaces. Following
Refs.~\cite{LK91,GK97} (cf. also Ref.\,\cite{HK2001}), we
implement the boundary conditions on
$S_\alpha$ using delta functions, leading to the partition
functions
\begin{subequations}
\begin{eqnarray}
{\cal Z}_{\text{D}} \, & = & \, \frac{1}{{\cal Z}_0} \int {\cal
D}\Phi \prod_{\alpha=1}^2 \prod_{X_{\alpha}}
\delta[\Phi(X_{\alpha})]
\exp(-S_{\text{E}}\{\Phi\}/\hbar)\, , \label{ZD} \\[2mm]
{\cal Z}_{\text{N}} \, & = & \, \frac{1}{{\cal Z}_0} \int {\cal
D}\Phi \prod_{\alpha=1}^2 \prod_{X_{\alpha}} \delta[\partial_n
\Phi(X_{\alpha})] \exp(-S_{\text{E}}\{\Phi\}/\hbar) \label{ZN} \,
\, ,
\end{eqnarray}
\end{subequations}
where ${\cal Z}_0$ is the partition function of the space without
plates. Note that in Eqs.\,(\ref{ZD}) and (\ref{ZN}) no gauge
fixing procedure is needed, since the relevant degrees of freedom
are expressed in terms of the fields $E_2$ and $B_2$ itself rather
than the vector potential $A_{\mu}$ in Eq.~(\ref{em-action}).  The
details of the calculation of the partition functions are left to
Appendix \ref{app_path}, and yield
\begin{equation}
\label{logZ} \ln {\cal Z}_{\text{D}} \,  =  \, - \, \frac{1}{2} \,
\mbox{tr} \ln \Gamma_{\text{D}},\quad \ln {\cal Z}_{\text{N}} \,
=  \, - \, \frac{1}{2} \, \mbox{tr} \ln \Gamma_{\text{N}} \, \, .
\end{equation}
The kernels $\Gamma_{\text{D}}$ and $\Gamma_{\text{N}}$ are given
by
\begin{subequations}
\label{gamma-kernel}
\begin{eqnarray}
[\Gamma_{\text{D}}]_{\alpha \beta}({\bf y},{\bf y}') \, & = & \,
[g_{\alpha}(y_1)]^{1/4} \, G[X_{\alpha}({\bf y})-X_{\beta}({\bf
y}')]
\, [g_{\beta}(y'_1)]^{1/4}\, , \label{AD} \\[2mm]
[\Gamma_{\text{N}}]_{\alpha \beta}({\bf y},{\bf y}') \, & = & \,
[g_{\alpha}(y_1)]^{1/4} \,
\partial_{n_\alpha(y_1)} \partial_{n_\beta(y'_1)}
G[X_{\alpha}({\bf y})-X_{\beta}({\bf y}')] \,
[g_{\beta}(y'_1)]^{1/4} \label{AN} \, \, ,
\end{eqnarray}
\end{subequations}
where $g_{\alpha}(y_1) = 1 + [h'_{\alpha}(y_1)]^2$
is the determinant of the induced metric, and
$n_\alpha(y_1)=(-1)^\alpha g_\alpha^{-1/2}(y_1)
[h'_\alpha(y_1),0,-1]$
is the normal vector to the surface $S_{\alpha}$, while
\begin{equation}
\label{G-free}
G(\underline{r})=\frac{1}{4\pi^2}\frac{1}{\underline{r}^2}
\end{equation}
is the {\it free} correlation function (Gaussian propagator)
corresponding to the Euclidean action $S_{\text{E}}$ in
Eq.\,(\ref{e-action}).  In the following, we denote position
vectors in the 4D Euclidean space by $\underline{r}=({\bf y},z)$.
We can then extract the Casimir energy
${\cal E}$ per unit area as
\begin{equation} \label{cale}
{\cal E}(H)  = E(H)  - \lim_{H \to \infty} E(H)\, ,
\end{equation}
with
\begin{equation} \label{e}
E(H) = - \frac{\hbar c}{A L} \left[ \ln {\cal Z}_{\text{D}} + \ln
{\cal Z}_{\text{N}} \right],
\end{equation}
where $A$ is the surface area of the plates, and $L$ denotes the
overall Euclidean length in time direction.

The above equations provide an {\em exact} result which yields the
Casimir energy for arbitrary static uniaxial deformations. These
equations can be used to evaluate the Casimir force by a recently
developed numerical approach \cite{E02}.  However, in order to obtain
a closed analytical expression for the partition functions in terms of
the kernels in Eqs.~(\ref{gamma-kernel}), here we resort to a
perturbative expansion with respect to the height profiles
$h_\alpha(y_1)$.  In fact, for this expansion to be valid, we have to
assume that the amplitude of the deformations sets the smallest
(geometric) length scale of the system. In what follows, perturbation
theory is carried out to second order in $h_\alpha(y_1)$,
separately for the two types of boundary conditions. However, within
second order in the height profile, the result is exact in the
sense that it correctly takes into account the many-body nature of the
Casimir interaction.

\subsection{Dirichlet boundary conditions}

Following Refs.~\cite{LK91,GK97}, we expand $\ln {\cal
Z}_{\text{D}}$ in a series $\ln {\cal Z}_{\text{D}}|_0 + \ln {\cal
Z}_{\text{D}}|_1 + \ln {\cal Z}_{\text{D}}|_2 + \ldots$, where the
subscript indicates the corresponding order in $h_\alpha$.  The
lowest order result is
\begin{equation} \label{h0D}
\ln {\cal Z}_{\text{D}}|_0 = \frac{AL}{H^3} \, \frac{\pi^2}{1440},
\end{equation}
corresponding to two flat plates. The first order result $\ln
{\cal
  Z}_{\text{D}}|_1$ vanishes since we assume, without loss of
generality, that the mean deformations are zero, $\int dy_1
h_{\alpha}(y_1) = 0$.  The {\em complete\/} second order
contribution is given by
\begin{eqnarray}
\label{zd} \ln {\cal Z}_{\text{D}}|_2 \, & = & \, - \, \frac{1}{4}
\int d^3 y \, \left\{ [h'_1(y_1)]^2 + [h'_2(y_1)]^2
\right\} \int \frac{d^3 p}{(2 \pi)^3} \times 1\\
& & \, + \, \frac{\pi^2}{240} \, \frac{1}{H^5} \, \int d^3 y \,
\left\{ [h_1(y_1)]^2 + [h_2(y_1)]^2
\right\} \nonumber\\
& & \, + \, \frac{1}{2} \int d^3 y \int d^3 y' \,
K_{\text{D}}(|{\bf y} - {\bf y}'|) \, \, \left\{ - \frac{1}{2} \,
[h_1(y_1) - h_1(y'_1)]^2 - \frac{1}{2} \, [h_2(y_1) - h_2(y'_1)]^2
\right\} \nonumber\\
& & \, - \, \frac{1}{2} \int d^3 y \int d^3 y' \,
Q_{\text{D}}(|{\bf y} - {\bf y}'|) \, \, [h_1(y_1) h_2(y'_1) +
h_2(y_1) h_1(y'_1)] \, \, . \nonumber
\end{eqnarray}
In the first term, which is further discussed below, $h'_\alpha = {\bf
  \nabla} h_\alpha$.  The kernels appearing above are given by
\cite{remark2}
\begin{subequations}
\label{kernel-d}
\begin{eqnarray}
K_{\text{D}}(y) \, & = & \, F_1(y) \, \partial_z^2 G(y,0) \, + \,
F_1(y) F_5(y) \, + \, F_3(y)^2 \, ,
\label{kd} \\[2mm]
Q_{\text{D}}(y) \, & = & \, F_4(y) \, \partial_z^2 G(y,H) \, + \,
F_4(y) F_6(y) \, + \, F_2(y)^2 \label{qd} \, \, ,
\end{eqnarray}
\end{subequations}
with the set of functions
\begin{subequations}
\label{fd}
\begin{eqnarray}
F_1(y) & = & \, \int \frac{d^3 p}{(2 \pi)^3} \, e^{i {\bf p} \cdot
{\bf y}}
\, \frac{G(p,0)}{{\cal N}(p,H)\, ,}\\[5mm]
F_2(y) & = & \, \int \frac{d^3 p}{(2 \pi)^3} \, e^{i {\bf p} \cdot
{\bf y}} \, \frac{G(p,0)}{{\cal N}(p,H)}
\, \, \partial_z G(p,H)\, , \\[5mm]
F_3(y) & = & \, \int \frac{d^3 p}{(2 \pi)^3} \, e^{i {\bf p} \cdot
{\bf y}} \, \frac{G(p,H)}{{\cal N}(p,H)}
\, \, \partial_z G(p,H) \\[5mm]
F_4(y) & = & \, \int \frac{d^3 p}{(2 \pi)^3} \, e^{i {\bf p} \cdot
{\bf y}}
\, \frac{G(p,H)}{{\cal N}(p,H)} \, ,\\[5mm]
F_5(y) & = & \, \int \frac{d^3 p}{(2 \pi)^3} \, e^{i {\bf p} \cdot
{\bf y}} \, \frac{G(p,0)}{{\cal N}(p,H)}
\, \, [\partial_z G(p,H)]^2\, , \\[5mm]
F_6(y) & = & \, \int \frac{d^3 p}{(2 \pi)^3} \, e^{i {\bf p} \cdot
{\bf y}} \, \frac{G(p,H)}{{\cal N}(p,H)} \, \, [\partial_z
G(p,H)]^2 \, \, ,
\end{eqnarray}
\end{subequations}
where $y = |{\bf y}|$, $p = |{\bf p}|$, $G(p,z) = \frac{1}{2 p}
e^{- p
  |z|}$ is the partially Fourier transformed free propagator of
Eq.~(\ref{G-free}), and ${\cal N}(p,H) = [G(p,0)]^2 - [G(p,H)]^2$.
The functions in Eqs.~(\ref{fd}) can be calculated explicitly
(see Appendix \ref{app_dirichlet}), leading to the simple result
\begin{subequations}
\begin{eqnarray} \label{kdshort}
K_{\text{D}}(y) \, & = & \, 2 K_{\text{D},\infty}(y) + K_{\text{D,reg}}(y)\, ,\\
Q_{\text{D}}(y) \, & = & \, \frac{\pi^2}{128} \, \frac{1}{H^6 y^2}
\, \frac{\sinh^2(s)}{\cosh^6(s)} \, \, ,
\end{eqnarray}
\end{subequations}
with
\begin{equation}
K_{\text{D},\infty}(y)=\frac{1}{2 \pi^4 y^8}, \quad
K_{\text{D,reg}}(y) = -\frac{1}{2 \pi^4 y^8} + \frac{\pi^2}{128}
\, \frac{1}{H^6 y^2} \, \frac{\cosh^2(s)}{\sinh^6(s)},
\end{equation}
where $s = \pi y / (2 H)$. The kernel $K_{\text{D}}$ has two
contributions of different origin.  In the limit $H \to \infty$,
corresponding to two decoupled surfaces, one has $K_{\text{D}} \to
2K_{\text{D},\infty}$, while $Q_{\text{D}}(y)$ vanishes. Thus the
part $K_{\text{D},\infty}$ describes a single surface. The two
($H$-independent) single surface contributions have to be
subtracted from the total kernel $K_{\text{D}}$ in order to
obtain the regularized kernel $K_{\text{D,reg}}$ which has to be
used in the calculation of the Casimir energy in Eq.~(\ref{cale}).
For
finite $H$, the kernel $K_{\text{D}}(y)$ actually has contributions
from both {\em outside\/} and {\em inside} the cavity, whereas
$K_{\text{D},\infty}$
comes from outside and the second term of $K_{\text{D,reg}}$ from
inside. The kernel $Q_{\text{D}}(y)$ has only contributions from
inside the cavity.

It is instructive to discuss the meaning of the contributions to
$\ln {\cal Z}_{\text{D}}|_2$ in Eq.~(\ref{zd}). The terms in the
first row are $H$-independent and formally divergent. They do not
contribute to the Casimir force between the surfaces but yield a
quantum electrodynamical increase of the surface tension of the
individual surfaces after introducing a suitable short-distance
cutoff \cite{Schwinger+78}. The necessity for a cutoff stems from
our continuum approach which breaks down on microscopic length
scales. The remaining terms in Eq.~(\ref{zd}) all contribute to
the Casimir force (with $K_{\text{D}}$ replaced by
$K_{\text{D,reg}}$). The local contributions in the second row are
half (due to TM modes only) of the individual surface (non-mixed)
terms which follow in second order of perturbation theory in
$h_\alpha$ from the pairwise summation approach, cf.~the second
term in Eq.~(\ref{pws}).  The third row in Eq.~(\ref{zd})
describes non-local individual surface contributions which are
missing in the pairwise summation approach. Finally, the last row
accounts for contributions due to the interference between the two
surface profiles. Obviously, it has a more complicated form than
the corresponding last term in the approximative pairwise
summation result in Eq.~(\ref{pws}).

\subsection{Neumann boundary conditions}

Expanding $\ln {\cal Z}_{\text{N}}$ in a series with respect to
$h_\alpha$ as before, the lowest order result is the same as for
the Dirichlet case,
\begin{equation} \label{h0N}
\ln {\cal Z}_{\text{N}}|_0 \, = \, \frac{AL}{H^3} \,
\frac{\pi^2}{1440} \, \, ,
\end{equation}
and the first order result $\ln {\cal Z}_{\text{N}}|_1$ again
vanishes. The complete second order result assumes a similar form
as for Dirichlet boundary conditions. We find
\begin{eqnarray}
\label{zn} \ln {\cal Z}_{\text{N}}|_{2}\, & = & \, + \,
\frac{1}{4} \int d^3 y \, \left\{ [h'_1(y_1)]^2 + [h'_2(y_1)]^2
\right\} \int \frac{d^3 p}{(2 \pi)^3}\times 1 \\
& & \, + \, \frac{\pi^2}{240} \, \frac{1}{H^5} \, \int d^3 y \,
\left\{ [h_1(y_1)]^2 + [h_2(y_1)]^2
\right\} \nonumber\\
& & \, + \, \frac{1}{2} \int d^3 y \int d^3 y' \, K_{\text{N}}
(|y_0 - y_0'|, |{\bf y}_{\parallel} - {\bf y}_{\parallel}'|)
\left\{- \frac{1}{2} \, [h_1(y_1) - h_1(y'_1)]^2 - \frac{1}{2} \,
[h_2(y_1) - h_2(y'_1)]^2
\right\} \nonumber \\
& & \, - \, \frac{1}{2} \int d^3 y \int d^3 y' \,
Q_{\text{N}}(|y_0 - y_0'|, |{\bf y}_{\parallel} - {\bf
y}_{\parallel}'|) \, \, [h_1(y_1) h_2(y'_1) + h_2(y_1) h_1(y'_1)]
\, \, . \nonumber
\end{eqnarray}
The kernels for Neumann boundary conditions assume a more
complicated form since the normal derivative breaks the
equivalence of space and time directions. The result reads
\begin{subequations}
\label{kernel-n}
\begin{eqnarray}
K_{\text{N}}(|y_0|, |{\bf y}_{\parallel}|) \, & = & \, {\cal
F}_1(y) \, \partial_z^2 g(y,0) \, + \, {\cal F}_1(y) {\cal F}_5(y)
\, + \, {\cal F}_3(y)^2
\label{kn}\\
& & + \, \partial_i \partial_j \, [ {\cal F}_1(y) \, \partial_i
\partial_j G(y,0) \, + \, {\cal F}_1(y) \, \partial_i \partial_j
{\cal F}_7(y) \, + \, \partial_i {\cal F}_9(y) \, \partial_j {\cal
F}_9(y) ]
\nonumber\\
& & + \, 2 \, \partial_i \, [ {\cal F}_1(y) \, \partial_i g(y,0)
\, + \, {\cal F}_1(y) \, \partial_i {\cal F}_{11}(y) \, + \, {\cal
F}_3(y) \, \partial_i {\cal F}_9(y) ] \, \, ,
\nonumber \\[4mm]
Q_{\text{N}}(|y_0|, |{\bf y}_{\parallel}|) \, & = & \, {\cal
F}_4(y) \, \partial_z^2 g(y,H) \, + \, {\cal F}_4(y) {\cal F}_6(y)
\, + \, {\cal F}_2(y)^2
\label{qn}\\
& & + \, \partial_i \partial_j \, [ {\cal F}_4(y) \, \partial_i
\partial_j G(y,H) \, + \, {\cal F}_4(y) \, \partial_i \partial_j
{\cal F}_8(y) \, + \, \partial_i {\cal F}_{10}(y) \, \partial_j
{\cal F}_{10}(y) ]
\nonumber\\
& & + \, 2 \, \partial_i \, [ {\cal F}_4(y) \, \partial_i g(y,H)
\, + \, {\cal F}_4(y) \, \partial_i {\cal F}_{12}(y) \, + \, {\cal
F}_2(y) \, \partial_i {\cal F}_{10}(y) ] \, \, , \nonumber
\end{eqnarray}
\end{subequations}
where summation over $i = 1,2$ and $j = 1,2$ is understood.  Note
that $\partial_i$ and $\partial_j$ act on the {\em spatial\/}
components ${\bf y}_{\parallel}$ of ${\bf y}$ only.  This is the
reason why the rotational symmetry within the $3$ dimensional
${\bf y}$ space is broken for Neumann boundary conditions.  For
translationally invariant profiles $h_\alpha(y_1)$ only terms with
$i=j=1$ contribute as can be seen by integration by parts.
However, as in the case of Dirichlet boundary conditions, the
above result is valid for any $h_\alpha({\bf y})$ but then can no
longer be interpreted as the contribution from TE modes to the
electrodynamic Casimir energy.

Here, we have introduced
\begin{equation} \label{propnew}
g(y,z) \, = \, \int \frac{d^3 p}{(2 \pi)^3} \, e^{i {\bf p} \cdot
{\bf y}} \, g(p,z) \, ,
\end{equation}
with $g(p,z) = \partial_z^2 G(p,z) = \frac{p}{2} e^{-p |z|}$, and
the functions ${\cal F}_j(y)$ defined as
\begin{subequations}
\label{fn}
\begin{eqnarray}
{\cal F}_1(y) & = & \, \int \frac{d^3 p}{(2 \pi)^3} \, e^{i {\bf
p} \cdot {\bf y}}
\, \frac{g(p,0)}{\eta(p,H)} \, ,\\[5mm]
{\cal F}_2(y) & = & \, \int \frac{d^3 p}{(2 \pi)^3} \, e^{i {\bf
p} \cdot {\bf y}} \, \frac{g(p,0)}{\eta(p,H)}
\, \, \partial_z g(p,H) \, ,\\[5mm]
{\cal F}_3(y) & = & \, \int \frac{d^3 p}{(2 \pi)^3} \, e^{i {\bf
p} \cdot {\bf y}} \, \frac{g(p,H)}{\eta(p,H)}
\, \, \partial_z g(p,H) \, ,\\[5mm]
{\cal F}_4(y) & = & \, \int \frac{d^3 p}{(2 \pi)^3} \, e^{i {\bf
p} \cdot {\bf y}}
\, \frac{g(p,H)}{\eta(p,H)} \, ,\\[5mm]
{\cal F}_5(y) & = & \, \int \frac{d^3 p}{(2 \pi)^3} \, e^{i {\bf
p} \cdot {\bf y}} \, \frac{g(p,0)}{\eta(p,H)}
\, \, [\partial_z g(p,H)]^2  \, ,\\[5mm]
{\cal F}_6(y) & = & \, \int \frac{d^3 p}{(2 \pi)^3} \, e^{i {\bf
p} \cdot {\bf y}} \, \frac{g(p,H)}{\eta(p,H)}
\, \, [\partial_z g(p,H)]^2 \, , \\[5mm]
{\cal F}_7(y) & = & \, \int \frac{d^3 p}{(2 \pi)^3} \, e^{i {\bf
p} \cdot {\bf y}} \, \frac{g(p,0)}{\eta(p,H)}
\, \, [\partial_z G(p,H)]^2 \, , \\[5mm]
{\cal F}_8(y) & = & \, \int \frac{d^3 p}{(2 \pi)^3} \, e^{i {\bf
p} \cdot {\bf y}} \, \frac{g(p,H)}{\eta(p,H)} \, \, [\partial_z
G(p,H)]^2 \, ,
\end{eqnarray}
\begin{eqnarray}
{\cal F}_9(y) & = & \, \int \frac{d^3 p}{(2 \pi)^3} \, e^{i {\bf
p} \cdot {\bf y}} \, \frac{g(p,H)}{\eta(p,H)}
\, \, \partial_z G(p,H)  \, ,\\[5mm]
{\cal F}_{10}(y) & = & \, \int \frac{d^3 p}{(2 \pi)^3} \, e^{i
{\bf p} \cdot {\bf y}} \, \frac{g(p,0)}{\eta(p,H)}
\, \, \partial_z G(p,H)  \, ,\\[5mm]
{\cal F}_{11}(y) & = & \, \int \frac{d^3 p}{(2 \pi)^3} \, e^{i
{\bf p} \cdot {\bf y}} \, \frac{g(p,0)}{\eta(p,H)}
\, \, \partial_z g(p,H) \, \partial_z G(p,H) \, , \\[5mm]
{\cal F}_{12}(y) \, & = & \int \frac{d^3 p}{(2 \pi)^3} \, e^{i
{\bf p} \cdot {\bf y}} \, \frac{g(p,H)}{\eta(p,H)} \, \,
\partial_z g(p,H) \, \partial_z G(p,H) \, ,
\end{eqnarray}
\end{subequations}
with $\eta(p,H) = [g(p,0)]^2 - [g(p,H)]^2$.  The explicit form of
these functions can be found in Appendix~\ref{app_neumann}.

The result in Eq.~(\ref{zn}) has the same type of contributions as
discussed for the Dirichlet case. Both Dirichlet and Neumann
case include `surface tension' contributions, but with opposite signs, and
identical local terms (second row in Eq.~(\ref{zn})). Since these
local terms are the only (non-mixed) contributions obtained by the
pairwise summation approach, the latter does not distinguish
between the two types of boundary conditions.
The main results of our general analysis of surface deformations are
contained in Eqs.~(\ref{zd}), (\ref{zn}). In section~\ref{section_sinus}
we apply these results to the important case
of modulated plates.

\section{Pairwise summation approximation}
\label{secpws}

The path integral approach may be compared with the commonly
used approximative method of pairwise summation (PWS). In the latter
approach, the Casimir energy ${\cal E}(H)$ for two {\em arbitrary}
shaped bodies of mean distance $H$ is obtained by the pairwise
summation of a two-body potential $U(r)$. In terms of the
deformation fields $h_\alpha$ this leads to
\begin{equation}
\label{pws-def}
{\cal E}(H)=\frac{1}{A}\int d^2 {\bf y}_{\|} \int d^2
{\bf
  y}'_{\|} \int_{H+h_2(y_1)}^\infty dz \int_{-\infty}^{h_1(y_1)} dz'\,\,
U\left[ \left(({\bf y}_\|-{\bf y}'_\|)^2 + (z-z')^2\right)^{1/2}\right] \,\, .
\end{equation}
In general, these integrals need to be computed numerically. However,
there are the following simplifications. If one of the plates is flat,
e.g., $h_1(y_1)=0$, the integrals can be performed explicitly, leading
to the simple result
\begin{equation}
\label{PWS-exact}
{\cal E}(H)= \frac{1}{A} \int d^2 {\bf y}_{\|} \,\, {\cal
E}_0\left[H+h_2(y_1)\right],
\end{equation}
where ${\cal E}_0$ is here the energy of two flat plates at distance
$H$, calculated from the same pair potential $U(r)$. Thus, in this
particular case the pairwise summation approximation is equivalent to
a geometrical average of the flat plate energy with locally varying
plate distance over the plate area.

For two deformed plates, the integrals in Eq.~(\ref{pws-def}) in
general can only be performed perturbatively in the height profile. To
do so, we follow the usual PWS approximation and assume a
'renormalized' retarded van der Waals potential \cite{Bordag+01,vdW},
\begin{equation}
U(r)=-\frac{\pi\hbar c}{24} r^{-7}.
\end{equation}  
The `renormalization factor' of the pair potential is chosen here such
that in the limit of two flat plates the exact Casimir result ${\cal E}_0$,
cf.~Eq.~(\ref{ce+}), is recovered. To second order in $h_\alpha$ one
obtains now
\begin{equation}
\label{pws} {\cal E}(H)=-\frac{\pi^2}{720} \frac{\hbar c}{H^3}
-\frac{\pi^2}{120}\frac{\hbar c}{H^5 A} \int d^2 {\bf y}_{\|}
\left[ h_1^2(y_1)+ h_2^2(y_1)\right] + \frac{\pi}{24} \frac{\hbar
c}{A} \int d^2 {\bf y}_{\|} \int d^2 {\bf y}'_{\|} \,
\frac{h_1(y_1+y'_1) h_2(y_1)}{(H^2+ {\bf y}_{\|}^{'2})^{7/2}} \,\, .
\end{equation}
For simple types of plate modulations, the integrals over the
deformation fields can be calculated easily. This will allow us a
direct assessment of the validity range of the pairwise summation
approach by comparing Eq.~(\ref{pws}) to the predictions of the
path integral technique.

\section{Modulated plates}
\label{section_sinus}

We now apply the results of Sec.~\ref{section_path} to static
uniaxial modulations of two parallel plates. In the first part we
focus on the {\em normal} Casimir force between a flat and a
corrugated plate. This force per unit area is defined as
\begin{equation}
F_{\rm n}=-\frac{\partial {\cal E}}{\partial H}\, ,
\end{equation}
in terms of the Casimir energy in Eq.~(\ref{cale}). However, in
most of the experiments, the flat plate is replaced by a spherical
lens with large radius $R \gg H$. In the latter case the normal
force can be obtained by using the Derjaguin approximation (DA)
(or proximity force rule) \cite{D34}, leading to
\begin{equation}
\label{F-DA}
F_{\rm DA,n}=2\pi R{\cal E}.
\end{equation}
Therefore, in the context of the normal Casimir force, we just
calculate ${\cal E}$ explicitly.

In the second part of this section we generalize our results for
two modulated plates with equal modulation length but with a phase
shift between them. Due to the broken translational symmetry,
there is now also a {\em lateral} force between the two plates
which arises solely from the cross-terms $\sim h_1 h_2$ in
Eqs.~(\ref{zd}), (\ref{zn}). If we denote the shift between the
two corrugations by the length $b$, the lateral force is obtained from
\begin{equation}
F_{\rm l}=-\frac{\partial {\cal E}}{\partial b}.
\end{equation}

\subsection{Normal force}
\label{sec4a}

%
\begin{figure}[t]
\begin{center}
\includegraphics[height=1.5in]{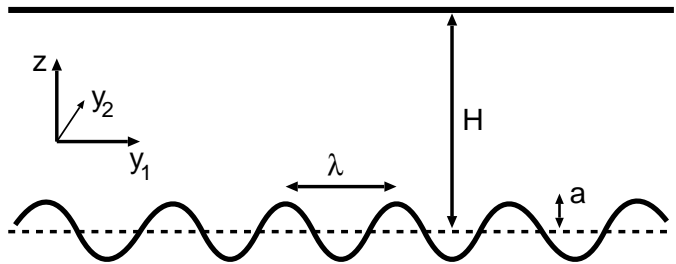}
\caption{The set-up used for calculating the Casimir energy between a
flat and a corrugated plate at mean separation $H$.}
\label{fig1}
\end{center}
\end{figure}
%

As a prototype of a corrugated surface, and to make contact with
recent experiments by Roy and Mohideen \cite{RM99}, we consider a
sinusoidally modulated plate along the $y_1$ direction, with
amplitude $a$, wavelength $\lambda$, and mean distance $H$ from
the flat plate (see Fig.~\ref{fig1}), i.e.,
\begin{equation} \label{corr}
h_1(y_1) \, = \, a \cos(2\pi y_1/\lambda) \, , \quad {\rm and}
\quad h_2(y_1) \, = 0 \, \, .
\end{equation}
For this particular deformation \cite{PGD2002}, only the
Fourier mode of wavelength $\lambda$ in the kernels $K_{\text{D}}(y)$
and $K_{\text{N}}(|y_0|, |{\bf y}_{\parallel}|)$ is probed. Thus, the
calculation of $\ln {\cal Z}_{\text{D}}$ and $\ln {\cal Z}_{\text{N}}$
via Eqs.~(\ref{zd}), (\ref{zn}) reduces to Fourier transforming the
kernels with respect to $y_1$. The corresponding expression for ${\cal
E}$ in Eq.~(\ref{cale}) can be written as
\begin{equation}
\label{ce} {\cal E} \, = \, {\cal E}_0 + {\cal E}_{\rm cf}
\, ,
\end{equation}
with
\begin{equation}
\label{ce+} {\cal E}_0 \, = \, - \, \frac{\pi^2}{720} \,
\frac{\hbar c}{H^3}, \quad {\cal E}_{\rm cf} \, = \, - \,
\frac{\hbar c a^2}{H^5} \left[ G_{\rm
TM}\left(\frac{H}{\lambda}\right)+ G_{\rm
TE}\left(\frac{H}{\lambda}\right)\! \right] \, + \, {\cal O}(a^3) \, ,
\end{equation}
where ${\cal E}_0$ is the energy of two flat plates and the index
cf of ${\cal E}_{\rm cf}$ stands for corrugated-flat geometry. 
The notation ${\cal O}(a^3)$ indicates that third and higher 
powers of $a/H$ and $a/\lambda$ are not considered here. 
The corrugation induced contributions to ${\cal E}_{\rm cf}$ from 
TM and TE modes at second order in $a$ are governed by the functions
\begin{subequations}
\label{gt}
\begin{eqnarray}
G_{\rm TM}(x) \, & = & \, \frac{\pi^2}{480} \, + \,
g_0(x) \label{gtm}\, ,\\[2mm]
G_{\rm TE}(x) \, & = & \, \frac{\pi^2}{480} \, + \, g_1(x) + x \,
g_2(x) + x^3 g_3(x) \, \, . \label{gte}
\end{eqnarray}
\end{subequations}
The first term $\pi^2/480$ in both equations corresponds to the
local contributions, cf.~the second row in Eqs.~(\ref{zd}),
(\ref{zn}). Since these are the only terms which are obtained
within a pairwise summation approach, the functions $g_m(x)$
represent non-trivial corrections which are neglected in the
pairwise summation scheme. These functions can be
calculated from the kernels $K_{\text{D}}(y)$ and
$K_{\text{N}}(|y_0|, |{\bf y}_{\parallel}|)$ by Fourier
transformation, leading to the expressions
\begin{subequations}
\label{g-fcts}
\begin{eqnarray}
\label{go} g_0(x) \, & = & \, - \, \frac{\pi^2}{480} \, + \,
\frac{\pi^3}{480} \, x \, +\, \frac{\pi^2}{128}
\int_{-\infty}^{\infty} ds \, \frac{\sin(4 x s)}{4 x s} \left[-
\frac{1}{s^6} + \frac{2}{15s^2} + \frac{\cosh^2(s)}{\sinh^6(s)}
\right]\, ,
\\[2mm]
g_1(x) \, & = & \, - \, \frac{\pi^2}{480} \, + \,
\frac{\pi^3}{1440} \, x
\label{g1} \\[2mm]
& & + \, \frac{\pi^2}{64} \int_{-\infty}^{\infty} ds \,
\frac{\sin(4 x s)}{4 x s} \bigg[- \frac{13}{4 s^6} - \frac{5}{3
s^4} + \frac{4}{45 s^2}
      + \frac{5}{2 s^5} \, \frac{\cosh(s)}{\sinh(s)}
      - \frac{3}{2 s^3} \, \frac{\cosh(s)}{\sinh^3(s)} \nonumber \\[1mm]
& & \qquad \qquad
      + \frac{1}{2 s} \, \frac{\cosh^3(s)}{\sinh^5(s)}
      + \frac{1}{s} \, \frac{\cosh(s)}{\sinh^5(s)}
      + \frac{1}{2} \, \frac{\cosh^4(s)}{\sinh^6(s)}
      + \frac{5}{4} \, \frac{\cosh^2(s)}{\sinh^6(s)}
      - \frac{1}{s^4} \, \frac{1}{\sinh^2(s)} \bigg] \nonumber \\[2mm]
& & - \, \frac{\pi^2}{64} \int_{-\infty}^{\infty} ds \, \cos(4 x
s) \bigg[\frac{1}{s^6} + \frac{1}{45 s^2} - \frac{2}{3 s^4}
      - \frac{1}{s^4} \, \frac{1}{\sinh^2(s)}
      + \frac{1}{s^5} \, \frac{\cosh(s)}{\sinh(s)}
      - \frac{1}{s^3} \, \frac{\cosh(s)}{\sinh^3(s)} \bigg] \, ,\nonumber \\[2mm]
g_2(x) \, & = & \, \frac{\pi^2}{64} \int_{-\infty}^{\infty} ds \,
\sin(4 x s) \bigg[\frac{2}{45 s} - \frac{5}{s^5}
      + \frac{1}{s^4} \, \frac{\cosh(s)}{\sinh(s)} \label{g2} \\[1mm]
& & \qquad \qquad
      + \frac{2}{s^3} \, \frac{1}{\sinh^2(s)}
      + \frac{1}{s^2} \, \frac{\cosh(s)}{\sinh^3(s)}
      + \frac{1}{s} \, \frac{\cosh^2(s)}{\sinh^4(s)} \bigg]\, , \nonumber \\[2mm]
g_3(x) \, & = & \, \frac{\pi}{32} \int_{-\infty}^{\infty} ds \,
\sin(4 x s) \bigg[- \frac{3}{s^3} - \frac{4}{3 s}
      + \frac{2}{s^2} \, \frac{\cosh(s)}{\sinh(s)}
      + \frac{1}{s} \, \frac{\cosh^2(s)}{\sinh^2(s)} \bigg] \, .\label{g3}
\end{eqnarray}
\end{subequations}
Before giving the explicit forms of these functions, let us
consider two limiting cases. For $x \to 0$, or $\lambda/H \to
\infty$, we have $g_0(x)$, $g_1(x) \to 0$, and $g_2(x)$, $g_3(x)$
converge to finite numbers, thus leaving in Eqs.~(\ref{gt}) only
the local contributions from the pairwise summation approach. In
the opposite limit $x \to \infty$, or $\lambda/H \to 0$, the
integrals in Eqs.~(\ref{g-fcts}) decay to zero, leading to
\begin{equation}
\label{g-fcts-large} g_0(x)=\frac{\pi^2}{480}\left( \pi x - 1 +
\frac{5\pi}{126}
  \frac{1}{x}\right)+{\cal O}(e^{-4\pi x}),\quad
g_1(x)=\frac{\pi^2}{480}\left( \frac{\pi}{3} x - 1 +
\frac{\pi}{18}
  \frac{1}{x}\right)+{\cal O}(e^{-4\pi x}),
\end{equation}
and both $g_2(x)$ and $g_3(x)$ are ${\cal O}(e^{-4\pi x})$. From
this result it is obvious that for $\lambda/H \to 0$ in
Eqs.~(\ref{gt}), the terms $\pi^2/480$ from the pairwise summation
approach are exactly canceled by corresponding terms of opposite
sign in the non-trivial corrections described by $g_0(x)$ and
$g_1(x)$. The most relevant contributions in this limit are now
provided by the first term in $g_0(x)$ and $g_1(x)$ in
Eq.~(\ref{g-fcts-large}), leading to the novel scaling behavior
$G_{\rm TM/TE}(x) \sim x + {\cal O}(1/x)$.

The integrals in Eqs.~(\ref{g-fcts}) can be carried out for
$\lambda>0$, or equivalently $x > 0$, by closing the integration
contour via a semi-circle at infinity in the upper half of the
complex $s$ plane, using the residue theorem \cite{Arfken}.  The
resulting sum of an infinite series of residues can be expressed
in terms of the polylogarithm function \cite{Erdelyi}
\begin{equation} \label{li}
{\rm Li}_n(z) \, = \, \sum_{\nu=1}^\infty \, \frac{z^\nu}{\nu^n}
\, \, ,
\end{equation}
leading to, with $u \equiv \exp(-4\pi x)$,
\begin{eqnarray}
G_{\rm TM}(x)&=&\frac{\pi^3x}{480}-\frac{\pi^2 x^4}{30} \ln(1-u) +
\frac{\pi}{1920 x} {\rm Li}_2(1-u) + \frac{\pi x^3}{24} {\rm
Li}_2(u)
+ \frac{x^2}{24} {\rm Li}_3(u) + \frac{x}{32\pi} {\rm Li}_4(u)\nonumber\\
&& + \frac{1}{64\pi^2} {\rm Li}_5(u)+\frac{1}{256\pi^3 x} \left(
{\rm Li}_6(u)-
\frac{\pi^6}{945}\right)\, , \label{tm} \\[5mm]
G_{\rm TE}(x)&=&\frac{\pi^3 x}{1440}-\frac{\pi^2 x^4}{30} \ln(1-u)
+\frac{\pi}{1920 x} {\rm Li}_2(1-u) -\frac{\pi
x}{48}\left(1+2x^2\right){\rm Li}_2(u)+\left(\frac{x^2}{48}-
\frac{1}{64}\right){\rm Li}_3(u)+\nonumber\\
&& + \frac{5x}{64\pi}{\rm Li}_4(u)+\frac{7}{128\pi^2} {\rm
Li}_5(u) +\frac{1}{256\pi^3x}\left(\frac{7}{2}{\rm Li}_6(u)-\pi^2
{\rm Li}_4(u) +\frac{\pi^6}{135}\right) \label{te} \, \, .
\end{eqnarray}
It should be noted that the appearance of the polylogarithm
function in quantum electrodynamics is also known from the fine
structure constant dependent corrections to the gyromagnetic ratio
of the electron \cite{Laporta+96}.

\begin{figure}[htb]
\begin{center}
\includegraphics[height=2in]{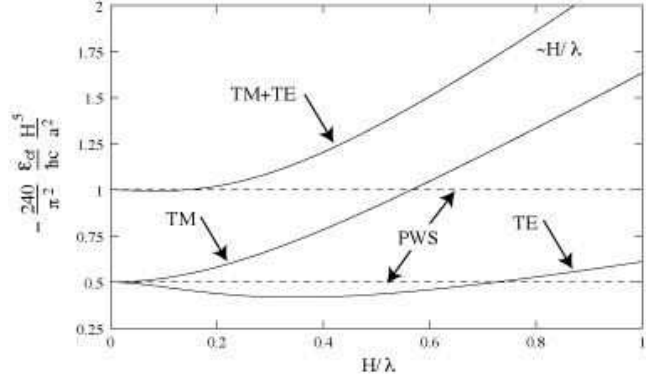}
\caption{Rescaled correction
  ${\cal E}_{\rm cf}$ to the Casimir energy due to the corrugation as
  given by the terms in square brackets of Eq.~(\ref{ce}) (upper
  curve).  The lower curves show the separate contributions from TM
  and TE modes.  The rescaling of ${\cal E}_{\rm cf}$ is chosen such
  that the corresponding prediction of the pairwise summation (PWS) 
  approximation [second term of Eq.\,(\ref{pws})] is a constant 
  (dashed lines).}
\label{fig2}
\end{center}
\end{figure}

Figure~\ref{fig2} displays separately the contributions from
$G_{\rm TM}$ and $G_{\rm TE}$ to the corrugation induced correction 
${\cal E}_{\rm cf}$ to the Casimir energy.  While $G_{\rm TM}(H/\lambda)$
is a monotonically increasing function of $H/\lambda$, $G_{\rm
  TE}(H/\lambda)$ displays a minimum for $H/\lambda \approx 0.3$.  

%
%
%
\begin{figure}[htb]
\begin{center}
\includegraphics[height=2.5in]{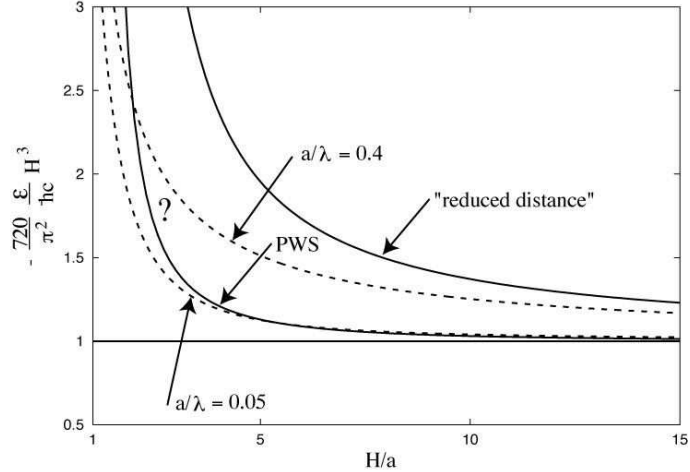}
\caption{Rescaled Casimir energy to second order in $a/H$ and
  $a/\lambda$ as given by Eq.~(\ref{ce}) for two fixed values of
  $a/\lambda$, shown as dashed curves.  The rescaling is chosen such
  that the Casimir energy of two flat plates becomes one (horizontal
  line).  The curve for $a/\lambda=0.05$ corresponds to the parameters
  used in the experiment of Ref.~\protect\cite{RM99}, where $H/a$
  varies between approximately 3 and 17. Note that the dashed lines
  are accurate predictions for the true Casimir energy in the limit $a
  \ll H,\lambda$ only, as indicated by the question mark (see text and
  the discussion in Sec.\,\ref{section_dis}). For comparison, the
  result of the pairwise summation (PWS) approximation
  [cf.~Eq.~(\ref{PWS-exact})] is shown. It agrees with the
  perturbative result in the limit $\lambda=\infty$ only.  In the
  opposite limit, $\lambda/a \to 0$, the energy can presumably be
  estimated from a reduced distance argument \cite{EHGK01,E02} (see
  Eq.~(\ref{reduced})).}
\label{fig3}
\end{center} 
\end{figure} 
%

Examining the limiting behaviors of Eq.~(\ref{ce}) is instructive.
In the limit $\lambda \gg H$, the functions $G_{\rm TM}$ and
$G_{\rm TE}$ approach constant values, and the Casimir energy
takes the $\lambda$-independent form
\begin{equation}
{\cal E}=-\frac{\hbar c}{H^3}\frac{\pi^2}{720}
\left(1+3\frac{a^2}{H^2} \right)
\, \, + \, {\cal O}(a^3) \, .
\end{equation}
Note that {\em only} in this case both wave types provide the same
contribution to the total energy and the result agrees with the
pairwise summation approximation (see Fig.~\ref{fig2}).  In the
opposite limit of $\lambda \ll H$, as demonstrated above, both
$G_{\rm TM}$ and $G_{\rm TE}$ grow linearly in $H/\lambda$.
Therefore, in this limit the correction to the Casimir energy
decays according to a {\em slower} power law in $H$, as
\begin{equation}
\label{large-H} {\cal E}=-\frac{\hbar c}{H^3}\frac{\pi^2}{720}
\left(1+2\pi\frac{a^2}{\lambda H}\right)
\, \, + \, {\cal O}(a^3) \, ,
\end{equation}
with an amplitude proportional to $1/\lambda$.  Note that this
behavior is completely missed by the pairwise summation approach
which always yields a $\lambda$ independent Casimir energy in the
presence of modulations on one plate.  

In the limit $\lambda \ll a, (H-a)$ we expect that the factor
multiplying $a/H$ in Eq.~(\ref{large-H}) saturates at a number of
order unity.  This result can be justified by noting that the most
relevant contributions to the force come from modes of wavelength of
order $H$. The corrugation also affects modes of wavelength of order
$\lambda$, but these modes contribute to the single plate energy only.
Thus, in the extreme limit $\lambda \ll a, (H-a)$, one has a clear
separation of the length scales $H$ and $\lambda$, and the modes
``see" flat plates at the reduced separation $H-a$, ($a>0$)
\cite{EHGK01}. More recently, an {\em exact} approach for calculating
the Casimir force has been developed which confirms the above argument
and yields for the case of TE modes (Dirichlet boundary conditions at
both plates) the exact result \cite{E02},
\begin{equation} \label{reduced}
{\cal E} = -\frac{\pi^2}{720}\frac{\hbar c}{(H-a)^3} \, \, ,
\quad \lambda \ll a, (H - a)
\, .
\end{equation}
This leads to a correction of the order $a/H$ (with prefactor 3)
after expansion in $a$. 

The above behavior of the correction ${\cal E}_{\rm cf}$ for small and
large $H/\lambda$ clarifies the limits of validity of previous results
in the literature. The upper dashed line in Fig.~\ref{fig2} corresponds 
to the PWS approximation (see Refs.\,\cite{Bordag+01,vdW} and 
Sec.\,\ref{secpws}). It is evident that this approximation
is accurate only for $H/\lambda \to 0$ (which in this limit is
equivalent to the Derjaguin method to any order in the amplitude $a$
\cite{D34}).  Already for $H/\lambda$ of order unity, the PWS
approximation breaks down.  The opposite limit,
$H/\lambda \to \infty$, corroborates the result reported in
Ref.~\cite{KNS87}, which is larger than the former by a factor of
$H/\lambda \gg 1$.  However, in experiments with lateral distortions
$\lambda$ of the order of $H$, none of the above limiting cases is
realized, which makes the present, more complete analysis necessary.

The use of a spherical tip, of large radius $R$, in experiments
\cite{RM99} causes some differences from the flat plate geometry used
in our calculations.  First, the positioning of the tip relative to
the modulations is important when $H$ and $\lambda$ are comparable,
but becomes insignificant in the proposed limit of $\lambda \ll H,R$.
Secondly, as long as $R\gg H, \lambda$ the curvature of the tip does
not lead to nontrivial corrections, and the force can be related to
the energy per surface area ${\cal E}$ in Eq.~(\ref{ce}) by the
proximity force rule $F=2\pi R {\cal E}$ in Eq.~(\ref{F-DA}).  
These formulas thus provide a specific recipe for evaluating 
the nontrivial shape dependences of the Casimir force in the 
experimental set-up.

The net Casimir energy ${\cal E}$ is shown in Fig.~\ref{fig3} for two
representative values of $a/\lambda$, including the parameters used in
the experiment of Ref.~\cite{RM99}. Note that the corrugation induced
correction leads to a larger energy ${\cal E}$, and hence the
corresponding (attractive) force $F=2\pi R {\cal E}$ is {\em
  enhanced}, at least to second order in $a/H$ and $a/\lambda$, which
becomes exact in the limit $a \ll H,\lambda$. This trend suggests, in
particular, that in the set-up of Fig.\,\ref{fig1} the force is always
attractive, although definite statements for values of $a/H$ and
$a/\lambda$ of order one can only be made by using non-perturbative
methods, as indicated by the question mark in Fig.\,\ref{fig3} (see
Ref.\,\cite{E02} and the discussion in
Sec.\,\ref{section_dis}). However, in the opposite limit, for which
the tips of the modulations of the lower plate in Fig.\,\ref{fig1}
almost touch the upper (flat) plate, i.e., $H - a \ll \lambda^2/a,a$,
the energy can be calculated {\em exactly\/} by using the Derjaguin
approximation for the individual tips of the modulations; this leads
to
\begin{equation} \label{opposite}
{\cal E} \, = \, - \, \frac{\pi^2 \sqrt{2}}{3840} \,
\frac{\hbar c}{a^{1/2} (H-a)^{5/2}} \, \, , \quad H - a \to 0 \, ,
\end{equation}
which corresponds to the result in Eq.~(\ref{PWS-exact}) after taking
$H-a \to 0$. The above result implies that at least for the particular
case of an uniaxial sinusoidal corrugation the corresponding force $F=2\pi R
{\cal E}$ is attractive when the surfaces almost touch.

\subsection{Lateral force}

\begin{figure}[t]
\begin{center}
\includegraphics[height=1.8in]{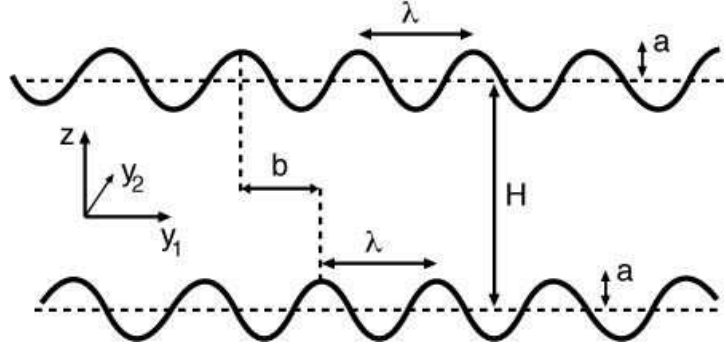}
\caption{Geometry used for calculating the lateral Casimir force
between two corrugated plates with lateral shift $b$. The equilibrium
position is at $b=\lambda/2$.}
\label{fig4}
\end{center}
\end{figure}

As a natural generalization of the geometry of the previous
section, we study the Casimir interaction between two sinusoidally
corrugated plates. For direct correspondence to recent
experiments \cite{Chen+02}, we consider the specific profiles
\begin{equation}
\label{corr-2} h_1(y_1) \, = \, a \cos(2\pi y_1/\lambda) \, ,
\quad {\rm and} \quad h_2(y_1) \, =  \, a \cos\left(2\pi
(y_1+b)/\lambda\right) \, \, ,
\end{equation}
which are shifted relative to each other by the length $b$ (see
Fig.~\ref{fig4}).

%
%

By inspection of the deformation dependent contributions to the
Casimir energy in Eqs.~(\ref{zd}) and (\ref{zn}), one obtains for the
total Casimir energy ${\cal E}$ of the corrugated-corrugated
geometry the relation
\begin{equation}
{\cal E} \, = \, {\cal E}_0 + 2 {\cal E}_{\rm cf} + {\cal E}_{\rm
cc},
\end{equation}
with ${\cal E}_0$ and ${\cal E}_{\rm cf}$ given in Eq.~(\ref{ce+}),
and the corrugation-corrugation interaction energy ${\cal E}_{\rm
cc}$ which can be calculated in terms of the kernels
$Q_{\text{D}}(y)$ and $Q_{\text{N}}(|y_0|, |{\bf y}_{\parallel}|)$
in Eqs.~(\ref{qd}) and (\ref{qn}). Besides oscillating contributions
to the normal Casimir force $F_{\rm n}(b)$ from ${\cal E}_{\rm cc}(b)$,
a {\em lateral} force
\begin{equation}
\label{f-lat} F_{\rm l}= - \frac{\partial {\cal E}_{\rm
cc}}{\partial b}\, ,
\end{equation}
is induced by the corrugation-corrugation interaction.
This lateral force is much better suited for experimental tests of the
influence of deformations, since there is no need for subtracting a
larger baseline force  (the contribution of flat plates)
as in the case of the normal force.

The calculation of the interaction energy ${\cal E}_{\rm cc}$ again
reduces for sinusoidally corrugated plates to Fourier transforming the
kernels $Q_{\text{D}}(y)$ and $Q_{\text{N}}(|y_0|, |{\bf
  y}_{\parallel}|)$. Separating the contributions from TM and TE
modes, we find
\begin{equation}
\label{Ecc} {\cal E}_{\rm cc} \, = \, \frac{\hbar c a^2}{H^5}
\cos\left(\frac{2\pi
 b}{\lambda}\right)
 \left[ J_{\rm TM}\left(\frac{H}{\lambda}\right)+
J_{\rm TE}\left(\frac{H}{\lambda}\right)\! \right] \, + \, {\cal O}(a^3) \, ,
\end{equation}
with
\begin{subequations}
\label{jt}
\begin{eqnarray}
J_{\rm TM}(x) \, & = & \, j_0(x) \, ,\label{jtm}\\[2mm]
J_{\rm TE}(x) \, & = & \, j_1(x) - x^2 \, j_2(x) + x^4 \, j_3(x)
\, \, , \label{jte}
\end{eqnarray}
\end{subequations}
and the functions $j_m(x)$ are given by
\begin{subequations}
\label{j-fcts}
\begin{eqnarray}
\label{j0} j_0(x) \, & = & \, \frac{\pi^2}{32}
\int_{-\infty}^{\infty} ds\, \frac{\sin(4 x s)}{4 x s}\,
\frac{\sinh^2(s)}{\cosh^6(s)} \, \, ,
\\[2mm]
j_1(x) \, & = & \, \frac{\pi^2}{32} \int_{-\infty}^{\infty} ds\,
\frac{\sin(4 x s)}{4 x s}\, \frac{\sinh^2(s)}{\cosh^6(s)} \left[
\frac{5}{2} - \sinh^2(s)\right] \, \, ,
\\[2mm]
j_2(x) \, & = & \, \frac{\pi^2}{4} \int_{-\infty}^{\infty} ds\,
\frac{\sin(4 x s)}{4 x s}\, \frac{\sinh^2(s)}{\cosh^4(s)}\, \, ,
\\[2mm]
j_3(x) \, & = & \, \frac{\pi^2}{2} \int_{-\infty}^{\infty} ds\,
\frac{\sin(4 x s)}{4 x s}\, \frac{\sinh^2(s)}{\cosh^2(s)}\, \, .
\end{eqnarray}
\end{subequations}
Before giving explicit forms for these integrals, it is
instructive to again consider their extreme limits.  For
$x=H/\lambda \to 0$, we find that both functions $J_{\rm TM}(x)$
and $J_{\rm TE}(x)$ tend to $\pi^2/240$. In the opposite limit
$x=H/\lambda \to \infty$ both functions decay exponentially fast
to zero so that the lateral force vanishes in this limit.  In
order to get the behavior in between these extremes, we have to
calculate the integrals in Eqs.~(\ref{j-fcts}). Using
the residue theorem, we finally obtain after summing over an
infinite series of residues by using the Lerch transcendent
\cite{Erdelyi}
\begin{equation}
\Phi(z,s,a)=\sum_{k=0}^\infty \frac{z^k}{(a+k)^s}\, ,
\end{equation}
the results (with $u\equiv \exp(-4\pi x)$),
\begin{subequations}
\label{j-pi}
\begin{eqnarray}
\label{j-tm} J_{\rm TM}(x) & = &
\frac{\pi^2}{120}\left(16x^4-1\right){\rm arctanh} (\sqrt{u})+
\sqrt{u}\left[\frac{\pi}{12}\left(x^3-\frac{1}{80x}\right)\Phi(u,2,\frac{1}{2})
+\frac{x^2}{12}\, \Phi(u,3,\frac{1}{2})\right. \nonumber\\
&& \left. + \frac{x}{16\pi}\, \Phi(u,4,\frac{1}{2})
+\frac{1}{32\pi^2} \, \Phi(u,5,\frac{1}{2}) +\frac{1}{128\pi^3 x}
\, \Phi(u,6,\frac{1}{2})\right]\, ,
\\[5mm]
\label{j-te} J_{\rm TE}(x) & = &
\frac{\pi^2}{120}\left(16x^4-1\right){\rm arctanh} (\sqrt{u})+
\sqrt{u}\left[-\frac{\pi}{12}\left(x^3+\frac{x}{2}+\frac{1}{80x}\right)
\Phi(u,2,\frac{1}{2})
+\frac{1}{24}\left(x^2-\frac{3}{4}\right)\Phi(u,3,\frac{1}{2})
\right. \nonumber\\
&& \left. + \frac{5}{32\pi} \left( x - \frac{1}{20x}\right)
\Phi(u,4,\frac{1}{2}) + \frac{7}{64\pi^2} \, \Phi(u,5,\frac{1}{2})
+\frac{7}{256\pi^3 x} \, \Phi(u,6,\frac{1}{2}) \right] \, \, .
\end{eqnarray}
\end{subequations}
This result can be compared to the pairwise summation approach by
considering the last term in Eq.~(\ref{pws}). For the surface
profiles considered here (cf.~Eq.~(\ref{corr-2})), this term
provides an interaction energy given by Eq.~(\ref{Ecc}) with the
sum $J_{\rm TM}(x)+J_{\rm TE}(x)$ replaced by the function
\begin{equation}
\label{j-pws} J_{\rm PWS}(x)=\frac{\pi^2}{360}\left(4\pi^2 x^2 +
6\pi x +3 \right)\sqrt{u}.
\end{equation}
The two results agree for $x=H/\lambda \to 0$, since $J_{\rm
  TM}(0)+J_{\rm TE}(0)=\pi^2/120=J_{\rm PWS}(0)$.
At the other extreme of $\lambda \ll H$, both $J_{\rm TM}(x)+J_{\rm
TE}(x)$ and $J_{\rm PWS}(x)$ decay exponentially fast but with
different $H/\lambda$ dependent coefficients. In particular, for large
$x=H/\lambda$, we get to leading order
\begin{equation}
J_{\rm TM}(x)+J_{\rm TE}(x) = \frac{4\pi^2}{15}\left( x^4 + {\cal
O}(x^2)\right) \sqrt{u} \, \quad \mbox{($x \to \infty$),}
\end{equation}
in contrast to the $\sim x^2$ behavior in Eq.~(\ref{j-pws}).

%
\begin{figure}[t]
\begin{center}
\includegraphics[height=2.2in]{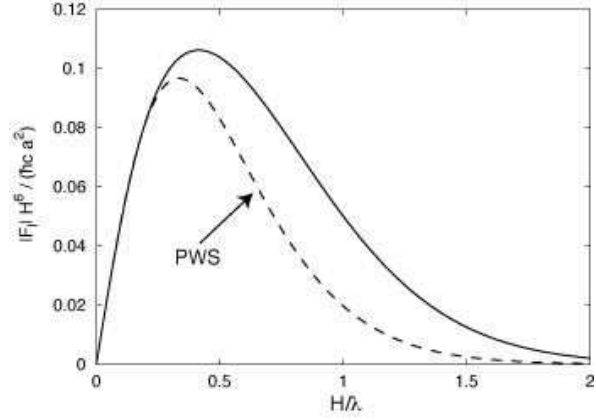}
\caption{Rescaled lateral Casimir force amplitude $|F_{\rm l}|$ as
  obtained from the path integral approach [Eq.~(\ref{j-pi})] (solid
  curve), and from the pairwise summation approach (PWS)
  [Eq.~(\ref{j-pws})] (dashed line).
The results hold to second order in $a$ 
(cf.~the discussion in Sec.\,\ref{section_dis}).}
\label{fig5}
\end{center}
\end{figure}
%

Since both $J_{\rm TM}(x)+J_{\rm TE}(x)$ and $J_{\rm PWS}(x)$ are
positive for all values of $x$, the equilibrium position of two
modulated surfaces is predicted at $b=\lambda/2$ in both approaches.
This corresponds to aligning the maxima and minima of the two corrugations
(cf.~Fig.~\ref{fig4}). The amplitude $|F_{\rm l}|$ of the lateral
force per unit area [Eq.~(\ref{f-lat})] as obtained from both
approaches is plotted in Fig.~\ref{fig5}. Interestingly, for fixed
$H$ there is an optimal modulation length $\lambda \approx 2.5 H$
at which the lateral force is largest. Our result shows that the
pairwise summation approximation is not justified beyond
$H/\lambda\approx 0.3$. For $H/\lambda$ of order one, the pairwise
summation approach has already a relative error of about $150\%$.
With increasing $H/\lambda$ this error grows monotonically.

\section{Discussion and outlook}
\label{section_dis}

We calculated normal and lateral Casimir forces between perfectly 
conducting modulated plates (Figs.\,\ref{fig1} and \ref{fig4}) 
by means of the path integral quantization method
(see Refs.~\cite{LK91,GK97}, and Sec.\,\ref{section_path}).
Based on the resulting {\em exact\/} expressions for the Casimir energy
(Eqs.\,(\ref{logZ}) - (\ref{e})), we performed a perturbative 
calculation to second order in the deformation parameter $a$
to obtain the results outlined in Sec.\,\ref{section_sinus}
and shown in Figs.\,\ref{fig2} and \ref{fig3} 
(normal force) and Fig.\,\ref{fig5} (lateral force).
These results are thus exact to second order in $a$,
and correctly take into account the many-body nature of the Casimir 
interaction, going beyond the commonly used pairwise summation (PWS) 
of van der Waals forces \cite{Bordag+01,vdW}. Our results
show significant deviations from PWS to second order in $a$.
However, for finite values of $a/\lambda$ and $a/H$, there will be 
corrections at higher orders in $a/\lambda$ and $a/H$;
in the present experiments, the sensitive range of $a/H$ is of the 
order of $0.2$ \cite{RM99,Chen+02}, while we suggest values for
$a/\lambda$ of order one to probe the nontrivial shape dependence
of the Casimir force (see below).
At present, it is not clear how relevant the perturbative 
results to second (or any finite) order in $a$ are for 
this range of the parameters $a/\lambda$ and $a/H$
(cf.~the related discussion in Sec.\,\ref{sec4a}).

To make further connection between our findings and the experimental
situations, corrections due to the finite conductivity of the plates,
finite temperature, and surface roughness should be taken into
account as well \cite{Bordag+01,KNS87,NSC,corr99,LR2000,corr00}.
These corrections introduce additional length scales into
the problem, which are in turn the plasma wavelength $\lambda_p$
of the plates (e.g., $\lambda_p \approx 100 \, \mbox{nm}$ for
aluminum \cite{RM99}), the thermal wavelength
$\lambda_T=\hbar c / k_B T$ ($\approx 8 \, \mu\mbox{m}$ at
$300^\circ \, \mbox{K}$), and the transverse correlation length
$\xi$ of the roughness (usually $\xi \approx 300 \, \mbox{nm}$
\cite{CAKBC2001}). While the effect of finite conductivity becomes
important for $H \lesssim \lambda_p$, finite temperature starts to
affect the result when $H \gtrsim \lambda_T$. The plasma and
thermal wavelengths thus provide lower and upper bounds for $H$,
respectively, such that our results for perfectly conducting
plates at zero temperature are valid for $\lambda, H \gg
\lambda_p$, and $H \ll \lambda_T$.

The importance of stochastic surface roughness can be deduced from
our calculations.  The relative corrections ${\cal E}_{\rm
  cf}/{\cal E}$ to the Casimir energy due to roughness of amplitude
$a$ and transverse correlation length $\xi$ should be of the form
(to second order in $a$)
$a^2/H^2$ for $\xi \gg H$, and $a^2/(\xi H)$ for $\xi \ll H$.  The
latter behavior is in accordance with Ref.~\cite{NSC}.  The
experimental case corresponds to neither extreme, making a more
complete analysis necessary.

Given the above mentioned limitations, as well as the technical
difficulties in achieving the desirable geometries in experiments,
it seems that it is difficult to conclusively establish the
nontrivial boundary dependence of the Casimir force.
Figure~\ref{fig3} shows that in the experiment of Roy and Mohideen
\cite{RM99}, the lengths $a$ and $\lambda$ are such that the
nontrivial dependence of the Casimir force on the boundary shape
is rather weak within the monitored range of $H/a$, and a pairwise
summation of two body forces is a possibly adequate approximation. Our
results suggest that a set-up with $\lambda$ of the order of $a$
is better suited for observing the nontrivial geometry dependence
predicted above.

In general, one expects that as long as the nontrivial features in
the geometry of the plates appear only as small perturbations to
the trivial flat-plate geometry, the corresponding many-body
effects of these features will be hard to measure. Considerably
larger effects could result, however, in patterned surfaces with
geometrical features that come close together across various parts
of the surfaces. In such circumstances a non-perturbative
calculation of the forces becomes necessary. Indeed it is most
desirable to find robust numerical schemes (possibly along the
lines of Ref.\,\cite{E02}) that can also incorporate 
the finite conductivity effects and surface roughness typical for 
experimental set-ups.

Finally, we note that in the set-up of Fig.~\ref{fig1}, nontrivial
shape dependencies appear as corrections to a larger Casimir
force.  For the purpose of experimental tests, it is much more
desirable to devise set-ups which directly probe differences,
without the need for subtracting a larger baseline force.  For
example, in an atomic force experiment, simultaneous scanning of a
flat and corrugated substrate would be desirable; while in the
torsion pendulum experiment, one can imagine suspending a
spherical lens equidistantly from two plates, one of which is
corrugated.

\section*{Acknowledgments}

We thank J.-P.~Bouchaud, S.~Dietrich, S.~G.~Johnson, U.~Mohideen,
M.~L.~Povinelli, and S.~Scheidl for useful discussions.  This work was
supported by the Deutsche Forschungsgemeinschaft through the Emmy
Noether grant No. EM70/2-1 (T.E.), through grant No.  HA3030/1-2
(A.H., at MIT), and by the National Science Foundation through grant
No. DMR-01-18213 (T.E. and M.K.).


\begin{appendix}

\section{Path integral formulation for partition functions}
\label{app_path}

We consider $N$ manifolds (objects) $\Omega_{\alpha}$ with $\alpha
= 1, \ldots, N$. Each point on the manifold $\Omega_{\alpha}$ is
represented by a vector $X_{\alpha}({\bf y}) =
(X_{\alpha}^{\mu}({\bf y}); \mu = 1, \ldots, d)$; a $D$
dimensional manifold $\Omega_{\alpha}$ embedded in $d$ dimensional
space is parameterized by ${\bf y} = (y_1, \ldots, y_D)$.

\subsection{Dirichlet boundary conditions}

The Dirichlet boundary condition $\Phi = 0$ on the manifolds,
can be enforced by the functional $\prod_{X_{\alpha}}
\delta[\Phi(X_{\alpha})]$ in Eq.\,(\ref{ZD}), which can be expressed in
terms of auxiliary fields $\Psi_{\alpha}(X_{\alpha})$ as
\cite{LK91,GK97}
\begin{equation} \label{auxiliary}
\prod_{X_{\alpha}} \delta[\Phi(X_{\alpha})] \, \equiv \, \int
{\cal D} \Psi_{\alpha}(X_{\alpha}) \, \exp\left[ i
\int\limits_{\Omega_{\alpha}} d X_{\alpha} \,
\Psi_{\alpha}(X_{\alpha}) \Phi(X_{\alpha}) \right] \, \, .
\end{equation}
The Gaussian integration over $\Phi$ in Eq.\,(\ref{ZD}) can then
be performed, resulting in
\begin{equation} \label{resulting}
{\cal Z}_{\text{D}} \, = \, \int \prod_{\alpha = 1}^{N} {\cal D}
\Psi_{\alpha}(X_{\alpha}) \, e^{-
\widetilde{S}_{\text{eff}}\{\Psi\}} \, \, .
\end{equation}
The effective action $\widetilde{S}_{\text{eff}}$ is given by
\begin{equation} \label{effective}
\widetilde{S}_{\text{eff}}\{\Psi\} \, =  \, \frac{1}{2} \,
\sum_{\alpha \beta} \int\limits_{\Omega_{\alpha}} d X_{\alpha}
\int\limits_{\Omega_{\beta}} d X_{\beta} \, \,
\Psi_{\alpha}(X_{\alpha}) \, G(X_{\alpha}, X_{\beta}) \,
\Psi_{\beta}(X_{\beta}) \, \, ,
\end{equation}
where $G(\underline{r}, \underline{r}')$ is the two-point
correlation function in unbounded bulk space. The functional
integration over the curved manifolds $\Omega_{\alpha}$ in
Eq.\,(\ref{resulting}) is facilitated by introducing the new
fields $\psi_{\alpha}({\bf y}) \equiv
\Psi_{\alpha}[X_{\alpha}({\bf y})]$. However, this transformation
requires some care regarding the integration measure
$\int_{\Omega_{\alpha}} d X_{\alpha}$ in Eq.\,(\ref{effective}), as
well as the functional measure $\int {\cal D}
\Psi_{\alpha}(X_{\alpha})$ in Eq.\,(\ref{resulting}). The result
is \cite{measure,HK2001}
\begin{equation} \label{resultingnew}
\int \prod_{\alpha} {\cal D} \Psi_{\alpha}(X_{\alpha}) \, e^{-
\widetilde{S}_{\text{eff}}\{\Psi\}} \, = \, \int \prod_{\alpha}
{\cal D} \phi_{\alpha}({\bf y}) \, e^{- S_{\text{eff}}\{\phi\}} \,
\, ,
\end{equation}
where the field $\phi_{\alpha}({\bf y}) \equiv [g_{\alpha}({\bf
y})]^{1/4} \psi_{\alpha}({\bf y})$ is given for each manifold
$\Omega_{\alpha}$ in terms of the determinant $g_{\alpha}({\bf
y})$ of its induced metric
\begin{equation} \label{metric}
g_{\alpha, ij}({\bf y}) \, = \, \sum_{\mu= 1}^{4} \frac{\partial
X_{\alpha}^{\mu}}{\partial y_i} \frac{\partial
X_{\alpha}^{\mu}}{\partial y_j} \, \, .
\end{equation}
The new effective action $S_{\text{eff}}$ is then given by
\begin{equation} \label{effectivenew}
S_{\text{eff}}\{\phi\} \, =  \, \frac{1}{2} \, \sum_{\alpha \beta}
\int d^D y \int d^D y' \, \phi_{\alpha}({\bf y}) \, \Gamma_{\alpha
\beta}({\bf y}, {\bf y}') \, \phi_{\beta}({\bf y}') \, \, ,
\end{equation}
where
\begin{equation} \label{A}
\Gamma_{\alpha \beta}({\bf y}, {\bf y}') \, = \, [g_{\alpha}({\bf
y})]^{1/4} \, G[X_{\alpha}({\bf y}), X_{\beta}({\bf y}')] \,
[g_{\beta}({\bf y}')]^{1/4} \, \, .
\end{equation}
The functional measure $\int {\cal D} \phi_{\alpha}({\bf y})$ on
the right hand side of Eq.\,(\ref{resultingnew}) is the one conventionally
used on a flat manifold (the local coordinate system). The
corresponding Gaussian integrations can thus be performed,
resulting in Eqs.\,(\ref{logZ}) and (\ref{AD}), with
$\Gamma_{\text{D}} \equiv \Gamma$ from Eq.\,(\ref{A}). Note that
in the present formulation, the trace and products of $\Gamma$ are
carried out by integrating ${\bf y}$ over the flat manifold of the
local coordinate system. Any dependence of ${\cal Z}_{\text{D}}$
on the metric $g_{\alpha, ij}({\bf y})$ is contained explicitly in
the definition of $\Gamma$ in Eq.\,(\ref{A}).

\subsection{Neumann boundary conditions}

For the Neumann boundary condition $\partial_n \Phi = 0$ on the
manifolds, the boundary condition enforcing functional
$\prod_{X_{\alpha}} \delta[\partial_n \Phi(X_{\alpha})]$ in
Eq.\,(\ref{ZN}) can again be expressed in terms of the auxiliary
fields $\Psi_{\alpha}(X_{\alpha})$ as
\begin{eqnarray}
\prod_{X_{\alpha}} \delta[\partial_n \Phi(X_{\alpha})] \, & \equiv
& \, \int {\cal D} \Psi_{\alpha}(X_{\alpha}) \, \exp\left[ i
\int\limits_{\Omega_{\alpha}} d X_{\alpha} \,
\Psi_{\alpha}(X_{\alpha}) \partial_n \Phi(X_{\alpha})
\right] \label{auxiliary_N} \\[2mm]
& = & \, \int {\cal D} \Psi_{\alpha}(X_{\alpha}) \, \exp\left[ -
\, i \int\limits_{\Omega_{\alpha}} d X_{\alpha} \, [\partial_n
\Psi_{\alpha}(X_{\alpha})] \, \Phi(X_{\alpha}) \right] \, \, ,
\end{eqnarray}
where the second line follows from an integration by parts. The
Gaussian integration over $\Phi$ in Eq.\,(\ref{ZN}) can then be
performed, resulting in
\begin{equation} \label{resulting_N}
{\cal Z}_{\text{N}} \, = \, \int \prod_{\alpha = 1}^{N} {\cal D}
\Psi_{\alpha}(X_{\alpha}) \, e^{-
\widetilde{S}_{\text{eff}}\{\Psi\}}
\end{equation}
with the effective action
\begin{eqnarray}
\widetilde{S}_{\text{eff}}\{\Psi\} \, & = & \, \frac{1}{2} \,
\sum_{\alpha \beta} \int\limits_{\Omega_{\alpha}} d X_{\alpha}
\int\limits_{\Omega_{\beta}} d X_{\beta} \, \, [\partial_n
\Psi_{\alpha}(X_{\alpha})] \, G(X_{\alpha}, X_{\beta}) \,
[\partial_n \Psi_{\beta}(X_{\beta})] \label{effective_N} \\[2mm]
& = & \, \frac{1}{2} \, \sum_{\alpha \beta}
\int\limits_{\Omega_{\alpha}} d X_{\alpha}
\int\limits_{\Omega_{\beta}} d X_{\beta} \, \,
\Psi_{\alpha}(X_{\alpha}) \, [\partial_{n_{\alpha}}
\partial_{n_{\beta}} G(X_{\alpha}, X_{\beta})] \,
\Psi_{\beta}(X_{\beta}) \nonumber \, \, .
\end{eqnarray}
Calculations along similar lines as in the previous paragraph then lead to
Eqs.\,(\ref{logZ}) and (\ref{AN}).

\section{Calculation of the kernels}

\subsection{Dirichlet boundary conditions}
\label{app_dirichlet}

The kernels for Dirichlet boundary conditions were defined in
Eqs.~(\ref{kernel-d}) in terms of the functions in
Eqs.~(\ref{fd}). The explicit form of these functions is given by
\begin{subequations}
\begin{eqnarray}
F_1(y) & = & \, -\frac{1}{\pi^2 y^4} \, - \, \frac{\pi}{8H^3y} \,
\frac{\cosh(s)}
{\sinh^3(s)}\, ,\\[5mm]
F_2(y) & = &
\, - \frac{\pi}{16H^3y} \, \frac{\sinh(s)}{\cosh^3(s)}\, ,\\[5mm]
F_3(y) & = &
\, 2\partial_z^2 G(y,0) \, - \, \frac{1}{2} F_1(y)\, ,\\[5mm]
F_4(y) & = & \, -2F_2(y)\, ,
\end{eqnarray}
\begin{eqnarray}
F_5(y) & = & \, - \frac{1}{2} F_3(y)\, ,\\[5mm]
F_6(y) & = & \, - \partial_z^2 G(y,H) \, - \, \frac{1}{2} F_2(y)\, ,
\end{eqnarray}
\end{subequations}
with $s=\pi y/(2H)$.

\subsection{Neumann boundary conditions}
\label{app_neumann}

For Neumann boundary conditions the functions appearing in the
kernels in Eqs.~(\ref{kernel-n}) and defined in Eqs.~(\ref{fn})
have the explicit forms
\begin{subequations}
\begin{eqnarray}
{\cal F}_1(y) & = & \, 2G(y,0) \, + \,
\frac{1}{4 \pi H y} \, \frac{\cosh(s)}{\sinh(s)} \, ,\\[5mm]
{\cal F}_2(y) & = &
\, - \frac{\pi}{16 H^3 y} \, \frac{\sinh(s)}{\cosh^3(s)}\, , \\[5mm]
{\cal F}_3(y) & = & \, g(y,0) \, + \,
\frac{\pi}{16 H^3 y} \, \frac{\cosh(s)}{\sinh^3(s)}\, , \\[5mm]
{\cal F}_4(y) & = & \, \frac{1}{4 \pi H y} \,
\frac{\sinh(s)}{\cosh(s)}\, ,\\[5mm]
{\cal F}_5(y) & = & \, - \, \frac{1}{2} \, \partial_z^2 g(y,0) \,
+ \, \frac{\pi^3}{32 H^5 y} \, \frac{\cosh^3(s)}{\sinh^5(s)}
\, + \, \frac{\pi^3}{16 H^5 y} \, \frac{\cosh(s)}{\sinh^5(s)}\, , \\[5mm]
{\cal F}_6(y) & = & \, - \, \partial_z^2 g(y,H) \, - \,
\frac{\pi^3}{32 H^5 y} \, \frac{\sinh^3(s)}{\cosh^5(s)}
\, + \, \frac{\pi^3}{16 H^5 y} \, \frac{\sinh(s)}{\cosh^5(s)}\, , \\[5mm]
{\cal F}_7(y) & = &
\, - G(y,0) \, + \, \frac{1}{4} {\cal F}_1(y) \, ,\\[5mm]
{\cal F}_8(y) & = & \, -G(y,H) \, + \,
\frac{1}{4} {\cal F}_4(y)\, , \\[5mm]
{\cal F}_9(y) & = &
\, - 2 {\cal F}_7(y)\, , \\[5mm]
{\cal F}_{10}(y) & = &
\, - \frac{1}{2} \, {\cal F}_4(y)\, , \\[5mm]
{\cal F}_{11}(y) & = &
\, - \frac{1}{2} \, {\cal F}_3(y)\, , \\[5mm]
{\cal F}_{12}(y) \, & = & \, -g(y,H) \, - \, \frac{1}{2} {\cal
F}_2(y) \, ,
\end{eqnarray}
\end{subequations}
with $g(y,z)=\partial_z^2 G(y,z)$.

\end{appendix}


\begin{thebibliography}{99}
\bibitem{remark2}The results for the Dirichlet kernels
$K$ and $Q$ quoted in Ref.\,\cite{GK97} deviate from
Eqs.\,(\ref{kd}) and (\ref{qd}) since the last terms in
each of the two lines of Eq.\,(15) in the second reference
\cite{GK97} are interchanged.
\end{thebibliography}
\end{document}